\documentclass[sigconf,nonacm]{acmart}
\setkeys{acmart.cls}{balance=false}

\usepackage{latexml}
\iflatexml
  \newcommand{\pdfauthororcid}[1]{}
  \newcommand{\authororcid}{}
\else
  \usepackage{orcidlink}
  \newcommand{\authororcid}{\,\orcidlink{0009-0005-8865-2090}}
  \makeatletter
  \newcommand{\pdfauthororcid}[1]{%
    \g@addto@macro\addresses{\g@addto@macro\@currentauthors{{\@authorfont\,\orcidlink{#1}}}}}
  \makeatother
\fi
\usepackage{stfloats}
\usepackage{balance}
\setcopyright{none}
\acmConference[Preprint]{Preprint}{September 2026}{Online}
\acmBooktitle{Preprint}
\acmDOI{}
\acmISBN{}
\acmYear{2026}
\copyrightyear{2026}
\title{Opacity Is Not Just Opacity}
\author{\texorpdfstring{Chunran Zhang\authororcid}{Chunran Zhang}}
\orcid{0009-0005-8865-2090}
\affiliation{%
  \department{School of Computing and Artificial Intelligence}
  \institution{Southwest Jiaotong University}
  \city{Chengdu}
  \country{China}}
\email{chronis@my.swjtu.edu.cn}
\author{Tianrui Li}
\orcid{0000-0001-7780-104X}
\authornote{Corresponding author.}
\pdfauthororcid{0000-0001-7780-104X}
\affiliation{%
  \department{School of Computing and Artificial Intelligence}
  \institution{Southwest Jiaotong University}
  \city{Chengdu}
  \country{China}}
\email{trli@swjtu.edu.cn}
\renewcommand{\shortauthors}{Chunran Zhang and Tianrui Li}
\begin{document}
\begin{abstract}
Web graphics travel with content across pages and themes, where changing backgrounds can require recoloring and maintenance. Opacity already makes a fixed object's appearance depend on its background, yet is usually understood only as how much the object obscures it. In fact, opacity controls the scaling of the object-background color difference; transparency is only one effect of this relationship. Zero places the output at the background and one at the source color, but difference scaling need not stop at either position. We retain the compositing expression and extend the coefficient domain from \([0,1]\) to the real numbers: negative values reverse the difference, whereas values above one expand it in the same direction. We focus on same-direction expansion for reusing Web graphics across backgrounds. Each object carries a fixed source color and coefficient, while the actual background determines the enhancement direction. Background-adaptive contrast enhancement thus becomes part of the object's compositing properties, reducing the design and maintenance of separate color variants. The implementation reuses the original equation without increasing the per-pixel arithmetic operation count within the same pipeline. Enumerating all 8-bit sRGB source colors on 16 predefined light and dark canvases, a fixed \(\alpha=1.1\) increases the contrast ratio in 99.8145\% of combinations. Without changing source colors, 4.8346\% of all combinations newly reach the \(3:1\) contrast threshold. Output validation and timing across three browsers demonstrate implementation in the same WebGL pipeline, with no sustained additional runtime observed.

\end{abstract}
\ccsdesc[500]{Computing methodologies~Rendering}
\keywords{alpha compositing, color contrast, background adaptation, web graphics}
\maketitle
\hypersetup{pdfauthor={Chunran Zhang and Tianrui Li}}
\setcounter{dbltopnumber}{1}

\begingroup
\setlength{\emergencystretch}{1em}
\section{Introduction}\label{introduction}

Web graphics travel with content: the same icon or graphic must work across pages and themes. When the background changes, its original colors may no longer provide sufficient contrast, requiring recoloring and maintenance of context-specific choices \citep{microsoftIcons, chameleon2025}. Can the object remain unchanged and adapt naturally to different backgrounds through its existing compositing properties, without additional color transformations or computation steps?

Opacity, an existing compositing property of the object, already makes a fixed object's appearance depend on the background. Even with both source color and opacity fixed, the same object can produce different displayed colors on different backgrounds. Yet opacity is usually understood only as how much the object obscures its background, with zero denoting full transparency and one full opacity \citep{w3cCompositing}.

Let \(C_b\) denote the opaque canvas color after the background layers have been composited, \(C_s\) the object's source color, and \(\alpha\) its opacity. The displayed result is

\[
C_o=\alpha C_s+(1-\alpha)C_b.
\]

Rewriting this equation as

\[
C_o-C_b=\alpha(C_s-C_b)
\]

shows that alpha actually controls the scaling of the color difference between object and background. The transparency described by opacity is the manifestation of this relationship within the conventional interval.

At zero, the output coincides with the background; at one, it reaches the source color. But difference scaling need not stop at either position.

We therefore retain the compositing expression and extend the coefficient domain from \([0,1]\) to the real numbers. For \(0\leq\alpha<1\), the difference contracts, producing transparency. Below zero, the difference is scaled in the reverse direction; above one, it expands in the same direction, moving the output beyond the source color and away from the background. Different values of the same compositing property can therefore blend the object into its background, reverse the difference, or enhance the distinction between them. Opacity describes the property's role within the conventional interval, whereas object-background difference scaling permits a broader domain. Separating these interpretations allows the same property to support effects beyond transparency. We focus on background-adaptive contrast enhancement through same-direction expansion. Figure~\ref{fig:method} illustrates conventional transparency and same-direction expansion.

\begin{figure*}[t]
\centering
\includegraphics[width=\linewidth]{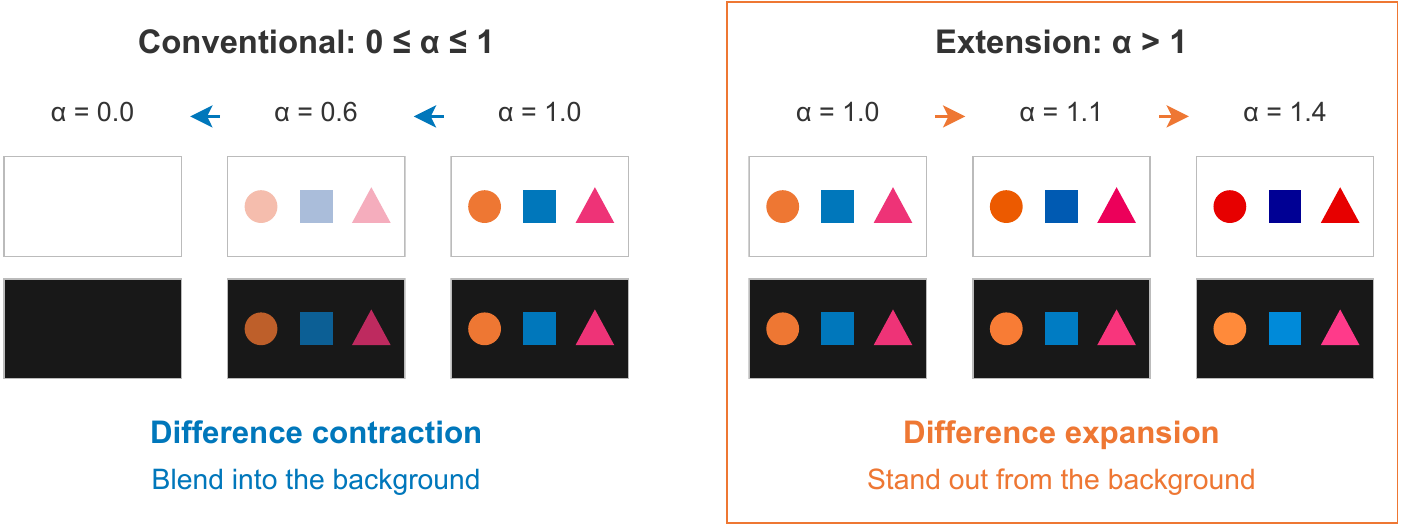}
\caption{The same source colors composited on light and dark canvases. The left panel shows contraction within the conventional interval; the outlined right panel shows same-direction expansion. At \(\alpha=1\), the source colors are preserved.}
\label{fig:method}
\Description{Contraction and expansion of fixed source colors on light and dark canvases.}
\end{figure*}

Within the same-direction expansion interval, the source color and coefficient remain fixed, while the actual background determines the expansion direction. The fixed configuration specifies the object's response to background changes. Its source color and coefficient travel with the graphic, while the displayed output is determined at its destination. Background-adaptive contrast enhancement thus becomes part of the object's compositing properties. In a compositing pipeline supporting extended coefficients, the application supplies the resolved background color at rendering time. The same icon or graphic can then carry a fixed configuration across pages, themes, and embedded canvases.

Specifying colors separately for each background requires storing those choices and maintaining them after design changes. Assigning adaptation to the compositing relationship concentrates design on the object's source color and coefficient: wherever the fixed configuration meets the target contrast ratio, no separate output color needs to be specified. Design changes can then be applied to the shared object definition, with outputs recomputed and checked across the target backgrounds.

This capability is obtained by extending the domain of the existing compositing coefficient. The source color, background color, and coefficient still enter the original equation. No separate color-search or transformation stage is introduced; within the same compositing pipeline supporting the extension, the per-pixel arithmetic operation count remains unchanged. A fixed configuration, background-adaptive contrast enhancement, and reuse of the computation follow from the same relationship.

We analyze how the coefficient controls object-background differences and which properties survive clipping. The numerical evaluation then examines the consequences for contrast ratio and color separation, keeping source colors and coefficients fixed across backgrounds. Predefined canvases establish the behavior on light and dark backgrounds, while coefficient sweeps and broader background tests characterize enhancement strength and its applicable range. Finally, browser output validation and timing compare conventional transparency, reverse scaling, and same-direction expansion within the same pipeline. This connects the unchanged expression to its rendered output and measured cost.

\endgroup
\section{Related Work}\label{related-work}

Contrast requirements for graphical objects reused across backgrounds are often addressed by changing their colors. Microsoft's icon guidelines allow separate assets for light and dark themes \citep{microsoftIcons}. Inverse Color Contrast Checker proposes replacement colors for Web text-background pairs \citep{sandnes2021}, while Iris repairs text and image contrast in Android apps \citep{iris2023}. Chameleon transforms light-mode visualization palettes for dark mode by optimizing luminance contrast, color consistency, and adjacent color differences \citep{chameleon2025}. These approaches store theme-specific color choices or compute replacement palettes, making adaptation a design and maintenance task or a separate repair process.

Palette generation also optimizes relationships among colors. Colorgorical balances discriminability and aesthetic preference \citep{colorgorical2017}; Palettailor jointly generates and assigns colors using the data distribution \citep{palettailor2021}. For transparent overlapped charts, Lu et al.~optimize colors, opacities, and rendering order to preserve category associations and segment discrimination \citep{lu2025}. These methods select visual encodings for a particular visualization; we fix source colors and coefficients to examine how their outputs vary across backgrounds.

Compositing already allows a fixed object to produce background-dependent outputs. W3C's Difference and Exclusion modes define different blending functions \citep{w3cCompositing}, while Kühne et al.~learn a hue-preserving operator from user preferences \citep{kuhne2012}. We retain the ordinary alpha compositing expression, reinterpret its coefficient through the object-background color difference, and extend its domain to the real numbers. Conventional transparency, reverse scaling, and same-direction expansion are expressed by different values of the same coefficient in the same expression.

Willis's Projective Alpha Colour develops a projective color representation and operational framework from color energy and carrying area. It explains premultiplied and non-premultiplied colors and discusses out-of-range alpha, source-over compositing, and filtering and illumination operations \citep{willis2006}. We start from the object-background difference relationship and ask how a compositing property can adapt fixed objects to different backgrounds. Neither the source color nor the coefficient needs to be reassigned for each background: background-adaptive contrast enhancement arises from the compositing relationship itself. We further analyze properties preserved after clipping under same-direction expansion and evaluate the contrast behavior of fixed configurations across backgrounds and their browser runtime.

Separate assets store background-specific color choices, while palette-generation and repair methods select colors for a context. Here, the background enters the original equation directly, and the object's existing coefficient specifies enhancement strength. Where the output meets the contrast target, the same object definition replaces a separately specified color. Within the same pipeline, this adaptation adds neither object parameters nor per-pixel arithmetic operations.

\section{Method}\label{method}

All color calculations and property analyses in this section use real-valued linear sRGB, with source and background colors in \([0,1]^3\).

\subsection{Compositing}\label{compositing}

Let \(C_s\) and \(C_b\) be the non-premultiplied source and background colors, with alpha values \(\alpha_s,\alpha_b\in[0,1]\). Standard source-over compositing gives \citep{porterDuff1984, w3cCompositing}

\begin{equation}
\begin{gathered}
c_o=\alpha_sC_s+(1-\alpha_s)\alpha_bC_b,
\\
\alpha_o=\alpha_s+(1-\alpha_s)\alpha_b.
\end{gathered}
\end{equation}

For \(\alpha_o>0\), the output color is \(C_o=c_o/\alpha_o\). Once background layers are composited onto the final opaque canvas, \(\alpha_b=1\), giving

\begin{equation}
C_o=\alpha_sC_s+(1-\alpha_s)C_b.
\end{equation}

\subsection{From Transparency to Signed Difference Scaling}\label{from-transparency-to-signed-difference-scaling}

With \(\alpha=\alpha_s\in[0,1]\), \(\alpha=0\) denotes full transparency, with output equal to the background; \(\alpha=1\) denotes full opacity, with output equal to the source color. Writing the output-background relationship as

\begin{equation}
C_o-C_b=\alpha(C_s-C_b)
\end{equation}

shows that alpha actually scales the color difference between object and background. Zero and one locate the background and source color, but difference scaling need not stop at either position.

We extend the compositing coefficient domain to \(\alpha\in\mathbb R\), retaining the original expression:

\begin{equation}
C_{\mathrm{raw}}=\alpha C_s+(1-\alpha)C_b.
\end{equation}

The sign determines the direction of the difference, and the magnitude determines its scale. Before clipping,

\begin{equation}
\begin{gathered}
C_{\mathrm{raw}}-C_b=\alpha(C_s-C_b),\\
\begin{cases}
\alpha<0 &: \text{reverse scaling},\\
0\leq\alpha\leq1 &: \text{conventional transparency},\\
\alpha>1 &: \text{same-direction expansion}.
\end{cases}
\end{gathered}
\end{equation}

In particular, \(\alpha=-1\) reverses the difference without changing its magnitude. If source and background colors coincide, every coefficient preserves the zero difference.

Relative luminance satisfies the same relationship:

\begin{equation}
Y_{\mathrm{raw}}-Y_b=\alpha(Y_s-Y_b).
\end{equation}

Thus, when source and background luminances differ and the output remains in gamut, \(\alpha>1\) increases the contrast ratio.

Out-of-gamut outputs require gamut mapping \citep{morovic2008}. We use per-channel clipping to \([0,1]\) to obtain the final output:

\begin{equation}
C_o=\operatorname{clip}_{[0,1]}(C_{\mathrm{raw}}).
\end{equation}

Within the conventional range \(\alpha\in[0,1]\), clipping leaves the result unchanged, preserving ordinary compositing in full.

\subsection{Fixed Objects under Background Changes}\label{fixed-objects-under-background-changes}

For a fixed source color \(C_s\) and coefficient \(\alpha\), the unclipped output responds to a background change as

\begin{equation}
\Delta C_{\mathrm{raw}}=(1-\alpha)\Delta C_b.
\end{equation}

For \(\alpha<0\), we have \(1-\alpha>1\): the output changes in the same direction as the background, with amplified magnitude. With source channels in \([0,1]\), a background channel of zero or one gives an unclipped output no greater than zero or no less than one, respectively; clipping preserves that background channel value. The final output therefore coincides with the background on pure black or pure white canvases.

For \(\alpha>1\), we have \(1-\alpha<0\), so the unclipped output changes in the opposite direction to the background. Fixed \(C_s\) and \(\alpha\) therefore specify an enhancement rule whose direction adapts to the actual background. The application supplies the resolved \(C_b\) at compositing time. With the same color space and clipping pipeline, conventional transparency, reverse scaling, and same-direction expansion execute the same expression, preserving the per-pixel arithmetic operation count. This adaptation uses the object's existing parameters and original computation, with no additional color-search or optimization stage.

\subsection{Enhancement Strength, Color Separation, and Background Differences}\label{sec:separation}\label{enhancement-strength-color-separation-and-background-differences}

The coefficient in a fixed configuration controls the direction and magnitude of difference scaling. For two objects using the same coefficient on the same background, their unclipped outputs satisfy

\begin{equation}
C_{\mathrm{raw},i}-C_{\mathrm{raw},j}
=\alpha(C_{s,i}-C_{s,j}).
\end{equation}

The common background term cancels, so the inter-object difference vector scales by the same coefficient, and its magnitude by \(|\alpha|\). For \(\alpha<0\), the difference reverses, contracting when \(|\alpha|<1\) and expanding when \(|\alpha|>1\). For \(\alpha>1\), the same coefficient expands both object-background and inter-object RGB differences before clipping.

Under either reverse or same-direction expansion, clipping can map distinct source colors to the same output. A channel that reaches a gamut boundary can no longer express further expansion in that direction. Distinct source colors can consequently share saturated channel values, reducing their separation in the final output. Within the same-direction expansion interval, reducing the coefficient towards one reduces clipping, controlling the tradeoff between background-adaptive contrast enhancement and inter-object color separation. The coefficient thus determines both the enhancement strength and color-separation cost of a fixed configuration.

This convergence occurs between objects; for \(\alpha\geq1\), output-background differences remain preserved. Including clipping, the final output satisfies

\begin{equation}
|C_{o,k}-C_{b,k}|
\geq |C_{s,k}-C_{b,k}|.
\end{equation}

Consequently, for any \(1\leq p\leq\infty\),

\begin{equation}
\|C_o-C_b\|_p\geq\|C_s-C_b\|_p.
\end{equation}

The RGB distance from output to background does not decrease. On black and white backgrounds, this per-channel guarantee also directly implies a nondecreasing contrast ratio. On black backgrounds, \(C_{o,k}\ge C_{s,k}\) in every channel, so relative luminance does not decrease. On white backgrounds, \(C_{o,k}\le C_{s,k}\) in every channel, so relative luminance does not increase. Both cases preserve or increase the contrast ratio against the background, including after clipping.

\begingroup
\setlength{\emergencystretch}{1em}
\section{Experiments}\label{experiments}

\subsection{Experimental Setup}\label{experimental-setup}

The experiments focus on background-adaptive contrast enhancement through same-direction expansion. We fix the object's source color and coefficient to assess contrast and color separation across backgrounds; runtime measurements cover conventional transparency, reverse scaling, and same-direction expansion.

The numerical evaluation enumerates all \(256^3\) 8-bit sRGB source colors \citep{srgb1996} on 16 black, white, near-black, near-white, and mildly tinted canvases. A \(65^3\) source grid is used to sweep \(\alpha\in[0,2]\) in steps of 0.025. A further evaluation combines \(17^3\) source colors with \(33^3\) background colors to broaden background coverage. These canvases and grids are synthetic test conditions, not a sample of color usage on the web.

All configurations use linear sRGB \citep{cssColor4} and per-channel clipping, with \(\alpha=1\) as the reference. We measure WCAG contrast-ratio changes \citep{wcag22}, using Difference and Exclusion as controls in the same color space \citep{w3cCompositing}. Color separation is measured by CIEDE2000 \citep{luo2001, sharma2005} and color collisions, defined as identical 8-bit outputs. We test all distinct pairs from Tol's seven-color Vibrant palette \citep{tol} and 65,536 distinct source-color pairs generated with a fixed seed. Appendix~\ref{app:reproducibility} details the metrics, backgrounds, and rendering.

\subsection{Background Contrast Enhancement with a Fixed Configuration}\label{background-contrast-enhancement-with-a-fixed-configuration}

The same source icons exhibit two roles of the compositing coefficient on light and dark canvases. Values \(0\leq\alpha<1\) blend the icons into the background, whereas \(\alpha>1\) expands their color differences from it (Figure~\ref{fig:icons}). Each configuration fixes the source colors and coefficient, with the actual background determining the displayed output.

\begin{figure}[t]
\centering
\includegraphics[width=\linewidth]{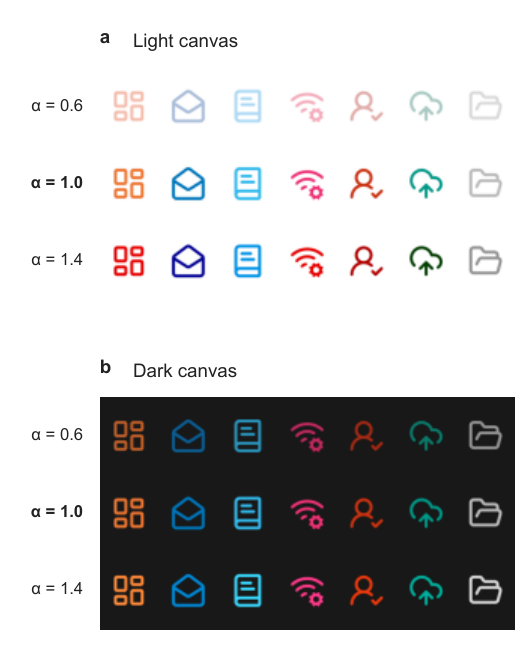}
\caption{Fixed icons on light and dark canvases. Rows correspond to \(\alpha=0.6,1.0,1.4\); each column retains the same icon and source color. These illustrations are rendered at 192 px.}
\label{fig:icons}
\Description{Seven icons at alpha 0.6, 1.0 and 1.4 on light and dark canvases.}
\end{figure}

Across all source colors and the 16 canvases, \(\alpha=1.1\) increases the contrast ratio in 99.8145\% of combinations, with a mean increase of 0.9783. Under the same conditions, the mean changes for Difference and Exclusion are \(-1.3050\) and \(-1.3104\), respectively.

At this coefficient, 4.8346\% of all source-canvas combinations newly reach the \(3:1\) threshold for meaningful graphical elements under WCAG non-text contrast \citep{wcagNonText}. These combinations reach the threshold while retaining their source colors, reducing the need for replacement colors to meet this contrast requirement.

Black and white canvases show no contrast-ratio decreases; other predefined canvases admit decreases. Their frequency increases on the broader background grid (Table~\ref{tab:backgrounds}).

\begin{table}[t]
\centering
\caption{Contrast-ratio changes under different background coverage. The predefined-canvas evaluation enumerates all 8-bit source colors; the broader evaluation uses \(17^3\) source colors and \(33^3\) backgrounds. Remaining combinations are unchanged within numerical tolerance.}
\label{tab:backgrounds}
\begin{tabular}{@{}crrrr@{}}
\toprule
& \multicolumn{2}{c}{Predefined (\%)} & \multicolumn{2}{c}{Broad grid (\%)} \\
\cmidrule(lr){2-3}\cmidrule(l){4-5}
$\alpha$ & Increase & Decrease & Increase & Decrease \\
\midrule
1.1 & 99.8145 & 0.1853 & 90.3672 & 9.3318 \\
1.4 & 99.6132 & 0.3867 & 86.7022 & 12.9968 \\
\bottomrule
\end{tabular}
\end{table}

On the broader grid, the maximum contrast-ratio losses are 0.4655 and 1.6394 at \(\alpha=1.1\) and \(1.4\), respectively. Among combinations with decreases, 99.34\% and 81.53\%, respectively, have losses no greater than 0.25. Loss denotes the absolute decrease in the contrast ratio.

Both coefficients incur their largest loss for a dark blue object on a yellow background: the source is sRGB \((0,0,0.8125)\) and the background is \((1,1,0)\). The original contrast ratio of 10.2773 falls to 9.8118 at \(\alpha=1.1\) and 8.6379 at \(\alpha=1.4\). Extrapolation drives the red and green channels below zero, where they are clipped, while the blue channel continues to increase. The object's luminance consequently moves closer to the background luminance. Nondecreasing per-channel background differences can therefore coexist with a decreasing contrast ratio.

\subsection{Enhancement Strength and Color Separation}\label{enhancement-strength-and-color-separation}

Background contrast enhancement must also account for color separation between objects. Within the tested same-direction expansion range, increasing the coefficient raises the mean contrast-ratio gain, but also makes distinct source colors more likely to produce identical outputs (Figure~\ref{fig:sweep}).

\begin{figure*}[t]
\centering
\includegraphics[width=\linewidth]{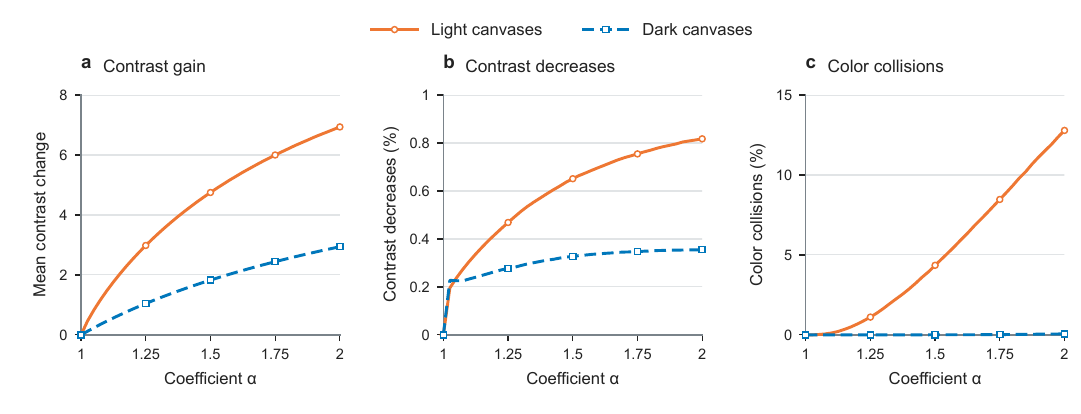}
\caption{Coefficient effects on enhancement and its costs. (a) Mean change in contrast ratio; (b) proportion of contrast-ratio decreases; (c) 8-bit color-collision rate. Each group weights eight canvases equally. Panels (a-b) use \(65^3\) source colors; panel (c) uses 65,536 distinct source-color pairs. All 41 measurements in \([1,2]\) are plotted.}
\label{fig:sweep}
\Description{Three coefficient-sweep panels show contrast gains, decreases and color collisions.}
\end{figure*}

Of the 336 Vibrant pair-background combinations, 32 have a smaller CIEDE2000 color difference at \(\alpha=1.1\), with no color collisions. At \(\alpha=1.4\), 88 combinations have smaller color differences, including eight collisions. Among the 1,048,576 random-pair-background combinations, collisions increase from 617 to 15,098.

Figure~\ref{fig:colors} shows the effects of clipping. As channels reach their limits, distinct source colors become less separated or coincide. Reducing the coefficient towards one reduces this convergence while reducing the extent of background contrast enhancement.

\begin{figure}[t]
\centering
\includegraphics[width=\linewidth]{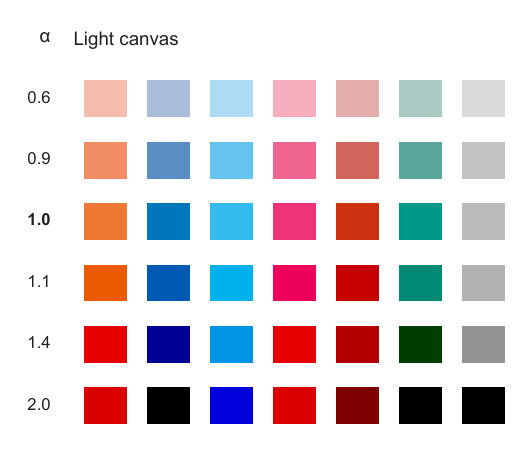}
\caption{Outputs for all seven Vibrant source colors on a white canvas. Columns fix the source color and rows vary the coefficient.}
\label{fig:colors}
\Description{All seven Vibrant colors at six coefficients on a white canvas.}
\end{figure}

The sweep quantifies the tradeoff between background contrast enhancement and color separation between objects. With source colors fixed, the coefficient specifies the enhancement strength and carries this design choice across canvases. Each background then determines the displayed output of the same fixed configuration.

Full icon arrays on near-source backgrounds are shown in Appendix~\ref{app:examples}.

\subsection{Web Implementation and Runtime}\label{web-implementation-and-runtime}

The WebGL implementation \citep{webgl} evaluates the original compositing expression in a custom fragment shader, with the real-valued coefficient supplied as a shader parameter. The object supplies the source color and coefficient, and the application-managed canvas supplies the background color. Geometric coverage is handled separately at antialiased edges. This keeps edge coverage independent of the coefficient's role in scaling the object-background color difference. Conventional transparency, reverse scaling, and same-direction expansion share the shader and rendering pipeline, differing only in the coefficient value.

Across 300 Lucide icons \citep{lucide}, three sizes, 16 canvases, and nine configurations, we generate 388,800 rendering instances across three browsers. For every pixel with nonzero coverage, the maximum channel error against the CPU reference is at most \(1/255\) in Chrome, Firefox, and WebKit. This validates implementation of the compositing relationship for the tested icons, sizes, and backgrounds.

Runtime measurements cover 161 coefficients over \(\alpha\in[-2,2]\) in steps of 0.025. The three browsers are measured sequentially on the same device. Each coefficient has 25 blocks of 20 consecutive frames, with coefficients randomly ordered within each round, yielding 241,500 timed frames in total (Figure~\ref{fig:runtime}). Appendix~\ref{app:rendering} specifies the hardware and software environments.

\begin{figure*}[t]
\centering
\includegraphics[width=\linewidth]{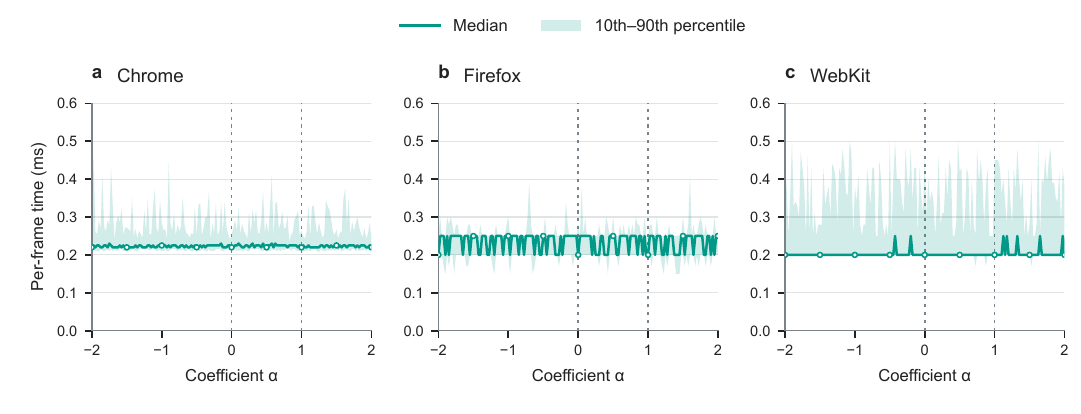}
\caption{Runtime across signed coefficient values. Lines show the median of the block means; bands show the empirical 10th-90th percentiles. The vertical dotted lines mark \(\alpha=0\) and \(1\). Each frame draws 300 icons and synchronously reads back one pixel. Firefox and WebKit block timings exhibit whole-millisecond steps.}
\label{fig:runtime}
\Description{Runtime over 161 coefficients from minus two to two in Chrome, Firefox and WebKit, with empirical percentile bands and boundaries at zero and one.}
\end{figure*}

Across coefficients, the median per-frame time computed from block means ranges from 0.220 to 0.230 ms in Chrome and from 0.200 to 0.250 ms in both Firefox and WebKit. Within the tested pipeline and coefficient range, neither reverse scaling nor same-direction expansion shows a sustained runtime increase relative to the conventional transparency interval.

\par
\endgroup
\balance
\begingroup
\section{Conclusion}\label{conclusion}

Interpreting alpha solely as opacity makes full transparency and full opacity the endpoints of its role. Yet the compositing equation scales the object-background color difference, with no algebraic restriction to coefficients between zero and one. We extend the coefficient domain from \([0,1]\) to the real numbers, allowing the same compositing property to express conventional transparency, reverse scaling, and same-direction expansion through different values of the original coefficient. Under same-direction expansion, the source color and coefficient remain fixed while the actual background determines the enhancement direction. Background-adaptive contrast enhancement thus becomes an inherent part of the object's compositing properties.

Across all 8-bit source colors and 16 predefined light and dark canvases, a fixed \(\alpha=1.1\) increases the contrast ratio in 99.8145\% of combinations. Without changing source colors, 4.8346\% of all combinations newly reach the \(3:1\) threshold. Background stress tests and coefficient sweeps characterize how clipping changes the displayed relationships. Under same-direction expansion, outputs can become less separated from one another while their RGB distances from the background remain nondecreasing. The contrast ratio can nevertheless decrease on some backgrounds. These behaviors distinguish what a fixed configuration preserves from what must be checked when selecting its coefficient.

Extending the coefficient domain reuses the object's source color, coefficient, and compositing expression, without a separate color-search or transformation stage. Within the same pipeline supporting extended coefficients, per-pixel arithmetic operation counts remain unchanged; cross-browser timing across all three intervals shows no sustained additional runtime. As graphics travel with Web content, their fixed source colors and coefficients specify their response at each destination; the actual backgrounds determine outputs through the same expression. Where the configuration meets the contrast target, no separate output color needs to be specified. Changing the shared source color or coefficient regenerates outputs through the same rule for checking across target backgrounds. Adaptation remains part of the object's compositing properties, reducing the maintenance of background-specific color choices. Opacity thus controls more than transparency: a broader relationship between object and background color differences. \emph{Opacity Is Not Just Opacity}.

\endgroup

\clearpage
\nobalance
\balance
\begin{acks}
The author(s) used OpenAI Codex for assistance with writing, coding, and visualization.

\end{acks}
\bibliographystyle{ACM-Reference-Format}
\begingroup
\interlinepenalty=10000
\setlength{\emergencystretch}{1em}
\bibliography{references}
\endgroup

\clearpage
\nobalance
\appendix
\interlinepenalty=10000
\section{Reproducibility Details}\label{app:reproducibility}\label{reproducibility-details}

\begin{figure*}[b]
\centering
\includegraphics[width=0.8\linewidth]{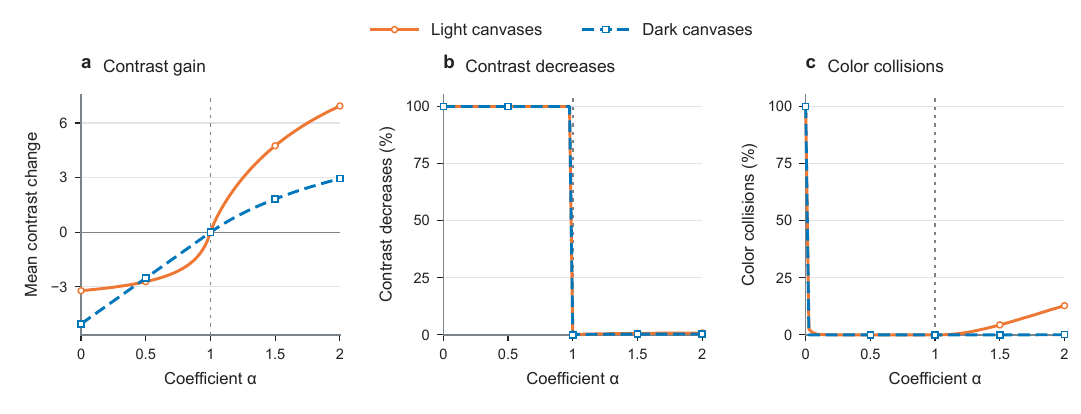}
\caption{Numerical coefficient sweep over \([0,2]\), using the same sources, backgrounds, and metrics as Figure~\ref{fig:sweep}. The dotted line marks \(\alpha=1\). Panels show mean contrast-ratio change, the proportion of contrast-ratio decreases, and the 8-bit color-collision rate.}
\label{fig:fullrange}
\Description{Numerical coefficient sweep from zero to two shows contraction and same-direction expansion, contrast decreases, and color collisions.}
\end{figure*}

Figure~\ref{fig:fullrange} shows the numerical sweep over \([0,2]\): contraction below one, source-color recovery at one, and expansion above one.

\subsection{Color Processing and Metrics}\label{color-processing-and-metrics}

Encoded sRGB inputs are decoded to linear sRGB before compositing \citep{cssColor4}. Relative luminance is \(Y(C)=0.2126C_R+0.7152C_G+0.0722C_B\), and the contrast ratio is \citep{wcag22}

\[
\operatorname{CR}(C_o,C_b)
=\frac{\max(Y_o,Y_b)+0.05}{\min(Y_o,Y_b)+0.05}.
\]

The reference is the source color on the same background. Calculations use float64, with contrast changes within \(10^{-10}\) treated as unchanged. Difference uses \(|C_s-C_b|\) and Exclusion uses \(C_s+C_b-2C_sC_b\), with componentwise operations in the same linear space.

Grids are uniform in encoded sRGB with both endpoints included, then decoded. The configuration records eight light and eight dark canvases.

CIEDE2000 uses CIELAB converted from linear sRGB with the D65 white point \citep{cie2018, cssColor4} and a decrease tolerance of \(10^{-9}\). Collisions use sRGB-encoded, 8-bit outputs rounded with ties to even. Random pairs use NumPy's default generator with seed 20260913. Equal-color cases and neutral sources are retained.

\subsection{Rendering and Timing}\label{app:rendering}\label{rendering-and-timing}

Hardware: MacBook Pro (Mac16,8), Apple M4 Pro, 14 CPU cores (10 performance, four efficiency), 20 GPU cores, and 24 GB unified memory. Output validation used macOS 26.6.2 (25G83); signed-coefficient timing used macOS 27.0 (26A428), both ARM64. User agents identify HeadlessChrome 152.0.0.0, Firefox 155.0, and WebKit 605.1.15 (Version 26.5). Chrome uses ANGLE/Metal.

The corpus uses the first 300 SVG paths from \texttt{lucide-static} 1.45.0, ordered by SHA-256 path hashes independently of rendering outcomes. Source colors cycle through the seven Vibrant colors. Browser validation uses 16, 24, and 48 px icons; the nine configurations are \(\alpha\in\{0.6,0.9,1,1.05,1.1,1.2,1.4\}\), Difference, and Exclusion.

The rasterized icon mask supplies geometric coverage \(q\in[0,1]\); the shader computes

\[
C_{\mathrm{pixel}}=q\,\operatorname{clip}_{[0,1]}
\bigl(C_b+\alpha(C_s-C_b)\bigr)+(1-q)C_b.
\]

Output is encoded to sRGB and written to an opaque canvas with output alpha fixed at one. Separating coverage from the extended coefficient preserves conventional antialiased edge coverage. The application supplies the already-composited background; the prototype does not read arbitrary DOM backgrounds or change native CSS opacity semantics.

Timing \citep{kalibera2013} uses 300 icons at 48 px on a \(1280\times960\) canvas, alternating warm light and cool dark backgrounds. Browsers run sequentially, each with two warm-up sweeps and 25 rounds over 161 coefficients in \([-2,2]\) at step 0.025, using identical seeded Fisher-Yates permutations per round. Each frame ends with synchronous one-pixel readback. \texttt{performance.now()} \citep{hrTime3} measures 20-frame blocks; division by 20 gives per-frame means. All 4,025 blocks per browser are retained, totaling 241,500 frames including draw submission and readback.

\subsection{Preservation of Background Distance}\label{preservation-of-background-distance}

For each channel, let \(s=C_{s,k}\), \(b=C_{b,k}\), \(s,b\in[0,1]\), and \(\alpha\geq1\). If \(s\geq b\), then \(b+\alpha(s-b)\geq s\); clipping leaves the output in \([s,1]\). If \(s\leq b\), clipping leaves it in \([0,s]\). In both cases, \(|C_{o,k}-b|\geq|s-b|\). Clipping therefore limits expansion at the gamut boundary without moving any channel back past its source value towards the background. Taking any \(\ell_p\) norm gives the inequality in Section~\ref{sec:separation}. This statement concerns real-valued RGB distances before output quantization.

\subsection{Coefficient Sweep and Visual Examples}\label{app:examples}\label{coefficient-sweep-and-visual-examples}

Figures~\ref{fig:gallery1} and~\ref{fig:gallery2} compare all 300 icons at \(\alpha=1\) and \(2\) on seven near-source backgrounds (Vibrant colors mixed with 10\% white in encoded sRGB) and a dark gray canvas. Source colors remain fixed; these examples illustrate appearance rather than recognition performance.

\begin{figure*}[p]
\centering
\includegraphics[width=0.90\linewidth]{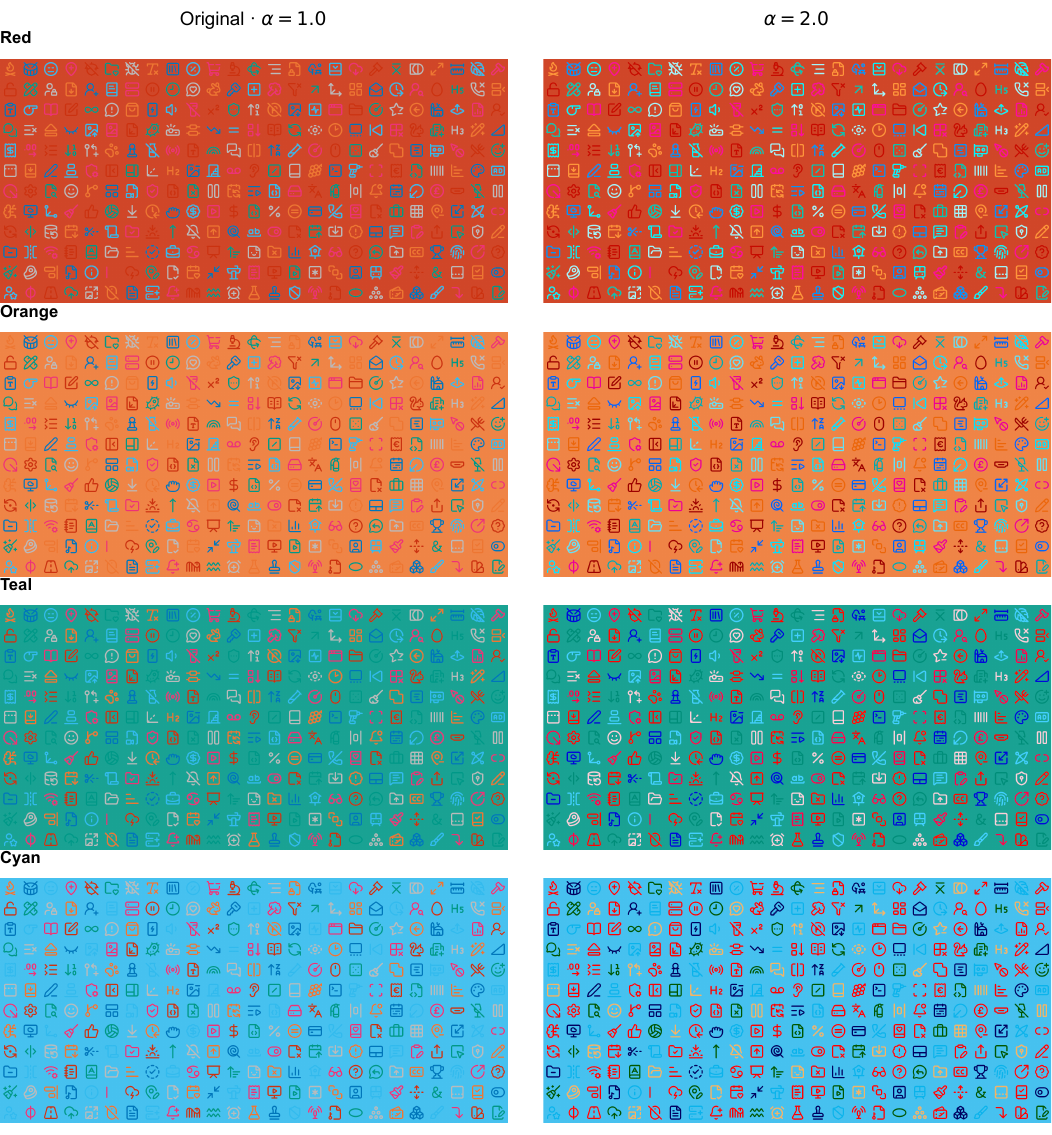}
\caption{Full icon arrays on near-source backgrounds, ordered from warm to cool hues. Each row compares the same 300 icons at \(\alpha=1\) (left) and \(2\) (right). Source colors, icon order, and dimensions are fixed across all panels.}
\label{fig:gallery1}
\Description{All 300 icons on red, orange, teal and cyan near-source backgrounds; alpha one on the left and two on the right.}
\end{figure*}

\begin{figure*}[p]
\centering
\includegraphics[width=0.90\linewidth]{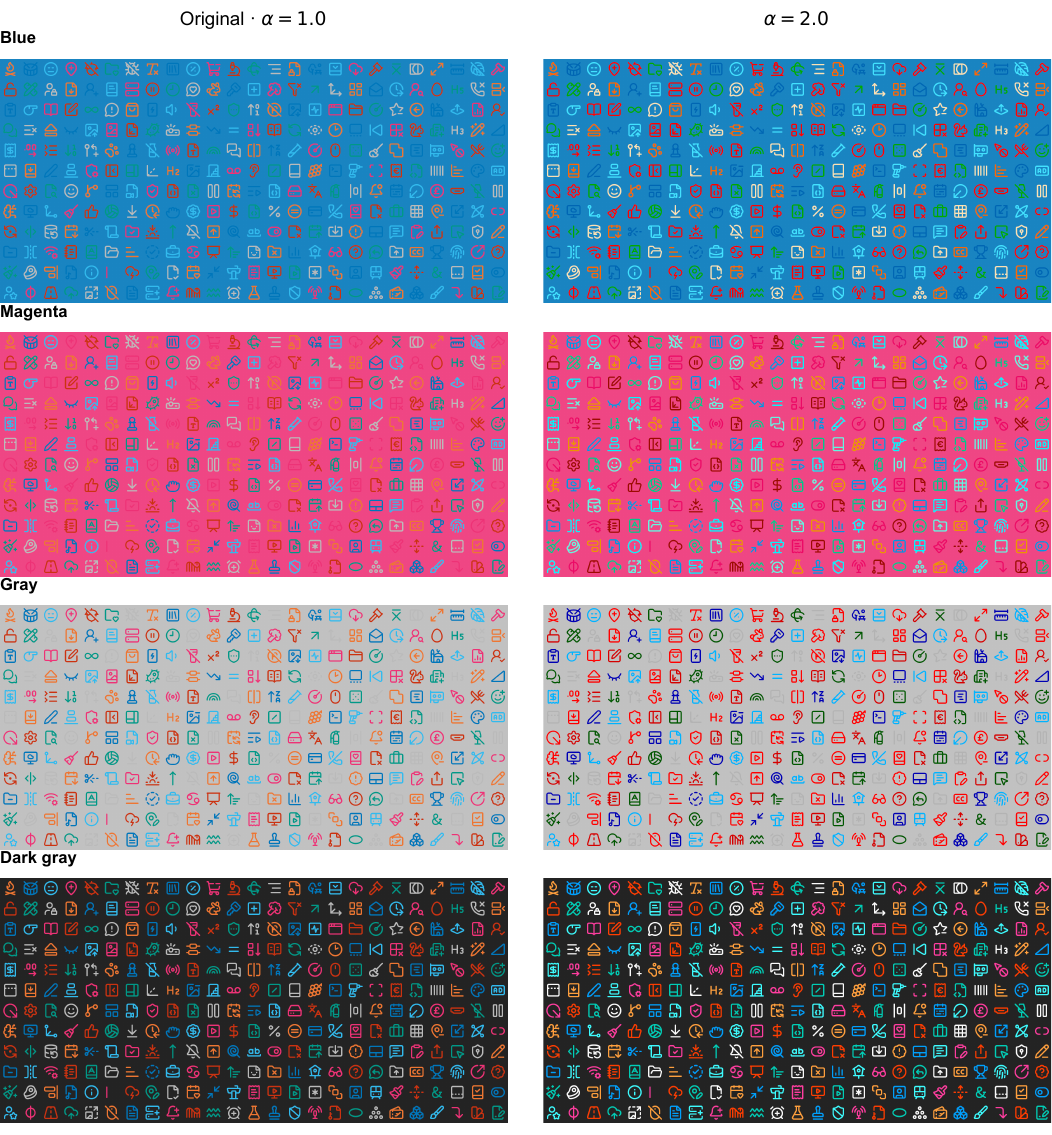}
\caption{Continuation of the background comparisons: blue, magenta, gray, and dark gray. The same source colors and icon order are retained; each row compares \(\alpha=1\) (left) with \(2\) (right).}
\label{fig:gallery2}
\Description{All 300 icons on blue, magenta, gray and dark gray backgrounds; alpha one on the left and two on the right.}
\end{figure*}

\end{document}